\documentstyle[aps,pre,epsf]{revtex}
\begin{document}
\twocolumn[\hsize\textwidth\columnwidth\hsize\csname @twocolumnfalse\endcsname

\title{Taming chaos by impurities in two-dimensional oscillator arrays}

\author{M.~Weiss, Tsampikos Kottos, and T.~Geisel\\
Max-Planck-Institut f\"ur Str\"omungsforschung, and \\
Institut f\"ur Nichtlineare Dynamik der Universit\"at G\"ottingen,\\
Bunsenstra\ss e 10, D-37073 G\"ottingen, Germany}
\maketitle

\begin{abstract}
The effect of impurities in a two-dimensional lattice of coupled nonlinear
chaotic oscillators and their ability to control the dynamical behavior of
the system are studied. We show that a single impurity can produce synchronized
spatio-temporal patterns, even though all oscillators and the impurity
are chaotic when uncoupled. When a small number of impurities is arranged in a way, 
that the lattice is divided 
into two disjoint parts, synchronization is enforced even for small coupling.
The synchronization is not affected as the size of the lattice increases, 
although the impurity concentration tends to zero.
\end{abstract}
]

\section{Introduction}
Coupled arrays of oscillators are studied extensively in many fields
of science because of their prevalence in nature. They are used as models for
coupled arrays of neurons~\cite{A89}, chemical reactions~\cite{chemreac}, coupled
lasers~\cite{tom} or Josephson junctions~\cite{UCM93}, charge-density-wave 
conductors~\cite{denswave}, crystal dislocations in metals~\cite{economou}, 
and proton conductivity in hydrogen-bonded chains~\cite{UMS98}. Various models and
coupling schemes have been proposed and analyzed previously~\cite{HPC95}.
A particular class are arrays of coupled oscillators, which exhibit chaotic 
motion when uncoupled. This class includes the forced Frenkel-Kontorova model~\cite{FM96}, 
which finds a straightforward physical realization in an array
of diffusively coupled Josephson junctions~\cite{gordon,pagano}, 
in which the applied current of each junction is modulated by a common frequency.
The possibility to obtain synchronized motion in such systems has been
investigated recently by Braiman et al. for the case of
one- (1D) and two-dimensional (2D) chaotic arrays of forced damped nonlinear
pendula~\cite{BLD95} and coupled Josephson junctions~\cite{braiman}. 
They observed the emergence of complex but frequency-locked spatio-temporal 
patterns, in which the chaotic behavior was completely suppressed, when a certain amount of 
{\it disorder} had been introduced by randomizing the lengths of the pendula.
In Ref.~\cite{GKKT98} the same phenomenon has been investigated from a completely 
different point of view: It was shown for 1D arrays of coupled chaotic pendula, 
that introducing a single impurity at a particular site is sufficient to 
lead to complete synchronization.

In this paper we study 2D arrays of coupled chaotic pendula. We ask for the minimal 
coupling as well as the influence of concentration and arrangement of impurities, needed 
to observe spatio-temporal 
patterns. Although for geometrical reasons one might expect that a single impurity cannot 
play the role it plays in 1D arrays, we find that it is able to tame the chaotic 
behavior of an arbitrarily large 2D array, provided that the coupling is strong enough.
A single impurity can produce synchronized spatio-temporal patterns, even though all 
oscillators and the impurity are chaotic when uncoupled.
In order to observe spatio-temporal patterns for smaller couplings, the
geometrical arrangement of the impurities is shown to play a crucial role.
Specifically we show that an impurity configuration that divides the lattice into
at least two disjoint parts is most appropriate for the creation of 
synchronized solutions. Such configurations will always produce patterns that locally
can be identified as lines, provided the coupling is above a threshold value.
The resulting spatio-temporal patterns are stable with respect to an increase of the
system size, indicating that an impurity concentration that tends to zero may 
suffice to lead to synchronization.


\section{Model}
We will focus our analysis on the model examined 
in Refs.~\cite{BLD95,GKKT98}:
\begin{eqnarray}\label{eqpend}
l_{n,m}^2 \ddot{\theta}_{n,m}+\gamma \dot{\theta }_{n,m}=
-gl_{n,m} \sin \theta _{n,m}+\tau ^\prime +\tau \sin \omega t \nonumber\\
+k(\theta _{n+1,m}+\theta _{n-1,m}-4\theta_{n,m} +\theta _{n,m+1}+
\theta _{n,m-1}),\,
\end{eqnarray}
where $n,m=1,2...N$. Thus, there is a damped, driven pendulum with unity mass and 
length~$l_{nm}$ on each site $(n,m)$ of the lattice, subject to an ac and a dc torque. 
The parameters used are the gravitational acceleration $g=1$, the dc torque $\tau^\prime=0.7155$, 
the ac torque $\tau =0.4$, the angular frequency $\omega =0.25$, and the damping $\gamma =0.75$. 
Neighbouring pendula are coupled via a discrete Laplacian, where~$k$ denotes the coupling 
strength. We have chosen free boundary conditions, 
i.e. $\theta_{0,m}=\theta_{1,m}$, $\theta_{N,m}=\theta_{N+1,m}, \theta_{n,0}= \theta_{n,1}$,
$\theta_{n,N}=\theta_{n,N+1} $ and used a fourth order Runge-Kutta routine 
with a time step $dt=0.01$ to numerically integrate Eq.~(\ref{eqpend}). We carefully checked 
that decreasing the time step to $dt=0.001$ did not alter our results.

A very convenient measure that allows a quick visualization of the average
global spatio-temporal behavior of the lattice, is the average velocity
\begin{equation}\label{2}
\sigma (jT)=\frac 1{N\cdot M}\sum_{n=1}^{N\cdot M}\dot{\theta }_n(jT)\,\, 
\end{equation}
at multiple times of the driving period~$T=1/\omega$.
Using this quantity~\cite{remark2} to obtain a bifurcation diagram not only can ascertain 
if chaotic or periodic behavior is obtained, but in addition helps to identify the 
maximum period of a pattern:
Computing~$\sigma(t)$ at each period~$t=jT$ of the driving, will lead to a periodic 
sequence $\sigma_1,...,\sigma_p,\sigma_1,...$, when transients have died out and a 
spatio-temporal pattern of periodicity~$p$ ('P$p$ attractor') has emerged. In practice we
inspect the last~$20$ values of $\sigma(jT)$ with $j=1,...170$, 
so that transients have died out. Thus, P20 attractors or attractors of larger periodicity 
are not recognized as such but rather appear as chaotic attractors.

Applying this strategy to an isolated pendulum, a bifurcation analysis with 
respect to the pendulum length~$l$ was performed in Ref.~\cite{GKKT98}. 
This approach revealed that each isolated pendulum is chaotic for values 
$l = 1\pm 0.002$ and that three more chaotic windows exist: 
Two narrow ones at $l\approx 0.84$, and $l\approx 0.52$, and a broad one for $l<0.35$. 
Thus we know whether a chosen pendulum with length~$l$ is chaotic or not. 
Furthermore, for the above chosen parameter values it is known, that
pendula with $l_{n,m}> 1$ and $l_{n,m}< 1$ show libration and rotation, respectively, apart
from the windows, where chaotic motion appears.


\section{Results and Discussion}
\begin{figure}
\begin{center}
\epsfxsize=8.2cm
\leavevmode
    \epsffile{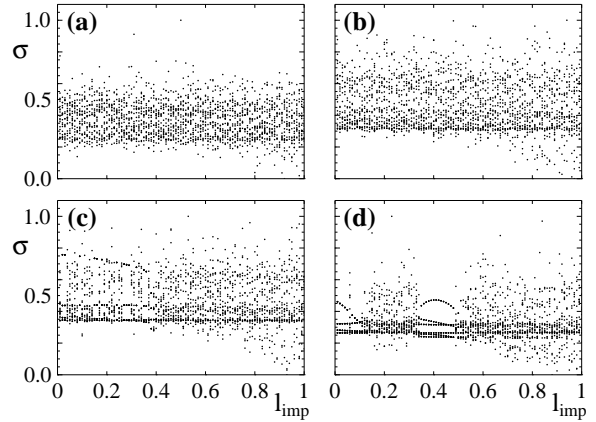}
\caption{Bifurcation diagram for a $50\times 50$ lattice of coupled 
pendula with length $l_{n,m}=1$. A single impurity with length $l_{\rm imp}$ is located at 
lattice site (25,25). For each $l_{\rm imp}$ the values of~$\sigma(151T),...,\sigma(170T)$
are shown, where~$T$ is the period of the driving. The coupling is {\bf (a)} $k=1$, 
{\bf (b)} $k=2$, {\bf (c)} $k=3$, and {\bf (d)} $k=5$. In (c) the first windows
of synchronization can be observed, which enlarge for bigger couplings (d).
}\label{fig:1}
\end{center}
\end{figure}

The simplest configuration of a 2D lattice described by Eq.~(\ref{eqpend}), 
is that of a single impurity in a sea of identical chaotic pendula with 
length~$l_{n,m}=1$. We fixed the lattice size to be~$50\times50$, with the
the impurity located at site $(25,25)$ and made a bifurcation analysis with respect 
to the impurity length $l_{\rm imp}$ for various values~$k$ of the coupling.
In Fig.~\ref{fig:1} the obtained bifurcation diagrams for $k=1,2,3,5$ 
are plotted versus~$l_{\rm imp}$. For convenience we normalized~$\sigma$ to take on 
values in the unit interval~$[0,1]$. Similarly to 1D arrays we find, 
that one impurity is able to organize the 2D arrays. However, our bifurcation analysis shows
that the coupling constant has to be larger than $k_{\rm cr}\simeq 3$, whereas
in the 1D case $k\geq 0.1$ was sufficient~\cite{GKKT98} to produce spatio-temporal
patterns. We would like to point out the appearance of a P4 attractor for 
$k=5,\,l_{\rm imp} < 0.15$ in Fig.~\ref{fig:1}d. Here an array of chaotic pendula is synchronized 
by a single impurity, which itself is chaotic, when isolated. 

\begin{figure}
\begin{center}
\epsfxsize=8.2cm
\leavevmode
    \epsffile{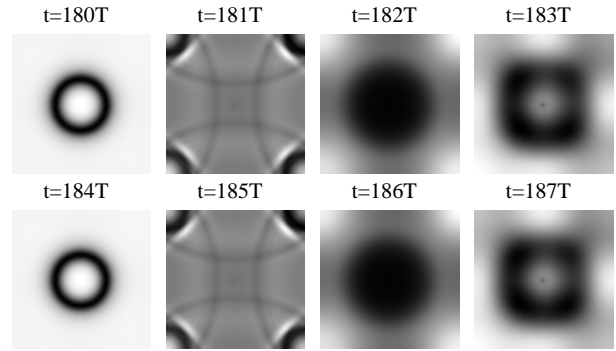}
\caption{Snapshots of $\dot{\theta }_{n,m}$ for a $64\times 64$ lattice 
with $l_{n,m}=1$, $l_{\rm imp}=l_{32,32}=0.1$ and $k=5$ at multiple times of the driving period~$T$. 
}\label{fig:2}
\end{center}
\end{figure}

In order to illustrate the occurence of 
this P4 pattern in a better way, we show in 
Fig.~\ref{fig:2} a typical gray scale plot of the velocities $\dot{\theta}_{n,m}$ 
as a function of the lattice coordinates $(n,m)$. The stroboscopic snapshots are 
taken with a time lag of one period of the driving~$T$; darker shading indicates higher 
velocity. In this example we used $64\times 64$ oscillators having $l_{n,m}=1$ and 
located an impurity of length $l_{imp}=0.1$ in the middle of the lattice at $(32,32)$. 
The coupling constant is $k=5$. After some transient (not shown) the 
synchronization is maintained and a P4 attractor is clearly visible.
We neglect edge effects except to remark that they are strictly confined to the
last few pendula near the boundaries. Moreover, they become very regular if the 
calculation is carried on for longer times. 

In many applications, however, one is interested in the weak coupling limit~\cite{WR90}. 
In this limit, our analysis shows that one impurity is not able to create spatio-temporal 
organization of a chaotic lattice. Therefore, it is meaningful to ask whether spatio-temporal 
patterns can emerge at all and under which conditions they may be obtained. 
To this end we increase the number of impurities introduced in the lattice and investigate the 
importance of their geometrical arrangement as well as their concentration.

\begin{figure}
\begin{center}
\epsfxsize=8.2cm
\leavevmode
    \epsffile{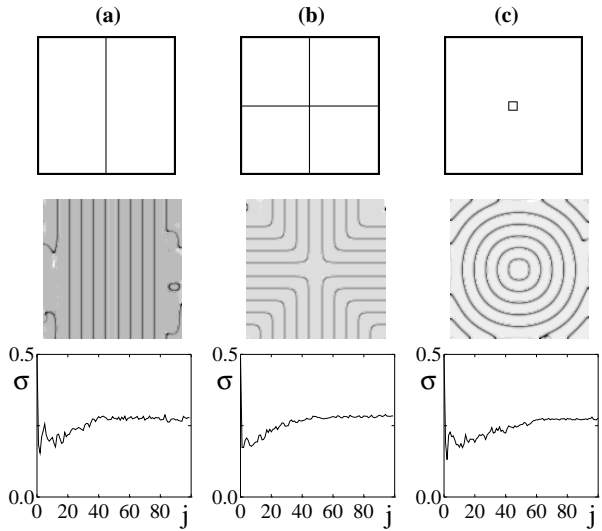}
\caption{Snapshots of $\dot{\theta }_{n,m}$ for a $128\times 128$ lattice 
with $l_{n,m}=1$, $l_{\rm imp}=0.7$ and $k=0.5$. The impurities are positioned
as indicated in the upper panel: {\bf (a)} on a line along $n=64$, {\bf (b)} on a 
cross along the lines $n=64$ and $m=64$, {\bf (c)} on a square with $(n,m)=(63,63)$ the 
lower left and $(66,66)$ the upper right corner. In the lower panel, the corresponding
$\sigma(jT)$ is shown.
}\label{fig:3}
\end{center}
\end{figure}

In Fig.~\ref{fig:3}a we report the P1 pattern emerging for a $128\times 128$ lattice 
with coupling constant $k=0.5$, when $128$ impurities were arranged along a 
line in the middle of the lattice, dividing it into two disjoint parts.
The same behavior could be observed even for smaller couplings.
We verified this for coupling constants as small as $k_{\rm cr}=0.1$.
Moreover, we found that increasing the size of the lattice to $N=256$ 
(the limit of our computational capability), but maintaining the line-like geometry
of impurities and the parameter-values, did not affect the formation of a pattern.
In all cases we observed synchronization to a P1 pattern after an initial transient.
Thus, increasing the size of the lattice will not affect the pattern formation 
although the percentage of the impurities will tend to zero in the limit of infinite systems. 
This indicates that the concentration of impurities is not of primary importance. 

To test the influence of the special arrangement of impurities, we considered a 
$128\times 128$ lattice with $k=0.5$ and $128$ impurities of length $l_{\rm imp}=0.7$
located at random positions of the lattice.
In all cases we have studied we obtained chaotic patterns. Even an increase 
of their concentration by more than a factor of~3 did not produce any spatio-temporal
pattern.

It is thus natural to ask wether a line of impurities is the only geometry 
that can produce spatio-temporal patterns for relatively small values of the coupling 
constant. In Fig.~\ref{fig:3}b,c we report the P1 attractors of a lattice of 
$128\times 128$ coupled pendula with lengths $l_{n,m}=1$, $l_{imp}=0.7$ and coupling $k=0.5$, 
where the impurities are arranged like a cross (Fig.~\ref{fig:3}b) or as a ring
(Fig.~\ref{fig:3}c). In both cases the observed pattern geometries consist locally of stripes
as was also observed for the 'line-geometry' in Fig.~\ref{fig:3}a. 

What is the common geometrical feature of the above impurity configurations that 
allow them to control the chaotic lattice? From the above analysis
we draw the conclusion, that it is sufficient for synchronization, that
the impurites divide the lattice into {\it at least} two disjoint 
subregions. In that way the critical coupling needed
to observe a spatio-temporal pattern is decreased by an order of magnitude. 
Reducing the coupling below $k_{\rm cr}\approx 0.1$, however, suppresses 
the formation of spatio-temporal patterns even for these lattices.
This finding is consistent with earlier investigations on 1D arrays, which revealed 
a critical coupling $k\approx0.1$, below which no pattern formation could be observed. 
Furthermore, the observed P1 pattern for the 'line geometry' (Fig.~\ref{fig:3}a)
is analogous to the 1D case, where the introduction of a single impurity caused the 
same topology of the array, i.e. a division into two disoint sets, and a similar pattern 
was observed~\cite{GKKT98}.
Thus, the results of the 1D case define the limiting $k$-value also for the 2D lattice.

\begin{figure}
\begin{center}
\epsfxsize=8.2cm
\leavevmode
    \epsffile{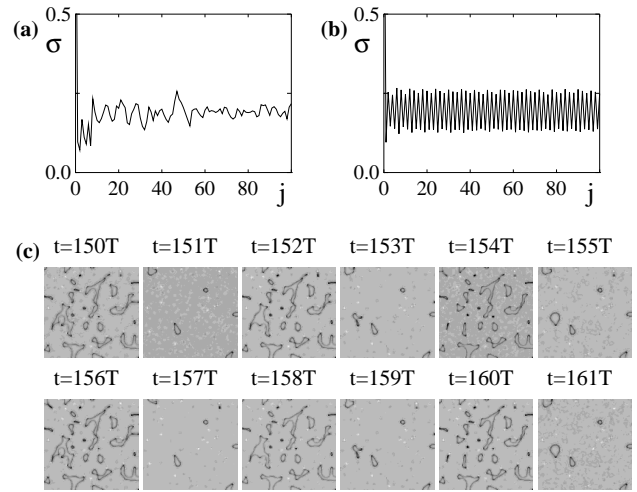}
\caption{{\bf (a)} Chaotic sequence $\sigma(jT)$ for a $128\times128$ lattice ($k=0.5$) with
$l_{n,m}\in[0.998,1.002]$. {\bf (b)} Periodic sequence $\sigma(jT)$ for a $128\times128$ lattice 
($k=0.5$) with $l_{n,m}\in[0.8,1.2]\backslash[0.998,1.002]$ revealing a P6 pattern.
{\bf (c)} Snapshots at multiple periods~$T$ of the driving confirm the existence of a P6 pattern
as predicted in (b).
}\label{fig:4}
\end{center}
\end{figure}

We finally examined the effect of disorder and the possibility
of obtaining self-organization or frequency locking. For a $128\times 128$ lattice
we randomly varied the lengths of the pendula but restricted the range of the disorder 
such, that each individual pendulum is chaotic, i.e. $l_{n,m}\in[0.998,1.002]$.
We found that the emerging pattern was always chaotic for a large number of different 
realizations of disorder and initial conditions.
The lacking synchronization can be observed in a better way by inspecting
$\sigma(jT)$, which does not show any periodicity (see Fig.~\ref{fig:4}a for an example).
If the range of disorder is increased to include also regular moving pendula, 
i.e. $l_{n,m}\in[0.8,1.2]$, self organization is possible in agreement 
with findings of Ref.~\cite{BLD95}. This can be understood by the observation that 
synchronization already occurs, when choosing the disorder from an interval of 
pendulum lengths associated with regular motion, i.e. $l_{n,m}\in[0.8,1.2]\backslash
[0.998,1.002]$.
A particular example for this is shown in Figs.~\ref{fig:4}b,c, where the 
occurence of a P6 pattern can be observed.
Thus, taking the lengths from the entire interval $l_{n,m}\in[0.8,1.2]$ on average 
yields only 1\% chaotic pendula, whose motion will be overdominated by the synchronizing 
regular ones, explaining the spatio-temporal pattern observed in Ref.~\cite{BLD95}. 

\section{Conclusion}

In conclusion we have demonstrated that a lattice of chaotic pendula can be
frequency locked into a spatio-temporal pattern by introducing impurities
in the lattice. In the strong coupling limit, a single impurity can tame
chaos. Decreasing the coupling constant requires more impurities in order
to observe self-organization. In this case the geometry of the impurity
configuration plays an important role. Our results suggest that if the
impurity configuration divides the lattice into at least two disjoint parts
then the coupling constant may be decreased without affecting the synchronization
of the chaotic array. Moreover, the induced spatio-temporal patterns are then unaffected
by the size of the lattice. Below the critical coupling~$k_{\rm cr}\approx0.1$ no synchronization
is observed. The value of this critical coupling is dictated by the minimum coupling which
leads to the formation of spatio-temporal patterns in the 1D case.

We are grateful to T.~Gavrielides, V.~Kovanis, A.~Politi, and G~Tsironis for helpful discussions.


\begin{thebibliography}{99}
%
\bibitem{A89} D.~Amid, {\it Modelling Brain Function}, Cambridge University
Press, Cambridge, UK 1989; J.~Hertz, A.~Krogh, R.~Palmer, {\it Introduction
to the theory of Neural Computation}, Addison-Wesley, Redwood City (1991).
%
\bibitem{chemreac}
A.~Arneodo, J.~Elezgaray, J.~Pearson, T.~Russo, Physica D {\bf 49} 141, (1991);
G.K.~Schenter, R.P.~McRae, B.C.~Garrett, J.Chem. Phys. {\bf 97}, 9116 (1992).
%
\bibitem{tom} G.~Kozyreff, A G.~Vladimirov, and P.~Mandel, Phys. Rev. Lett. {\bf 85}, 
3809 (2000); H.G.~Winful and L.~Rahman, Phys. Rev. Lett. {\bf 65}, 1575 (1990); 
J.~Terry, K.S.~Thornburg, A.D.J.~DeShazer, G.D.~VanWiggeren, S.~Zhu, and P.~Ashwin, 
Phys. Rev. E {\bf 59}, 4036 (1999); A.~Hohl, A.~Gavrielides, T.~Erneux, V.~Kovanis, 
Phys. Rev. Lett. {\bf 78}, 4745 (1997).
%
\bibitem{UCM93} A.V.~Ustinov, M.~Cirillo, and B.~Malomed, Phys. Rev. B
{\bf 47}, 8357 (1993); K.~Wiesenfeld, P.~Colet, and S.~Strongatz, Phys. Rev.
Lett. {\bf 76}, 404 (1996).
%
\bibitem{denswave}
S.H.~Strogatz-SH, C.M.~Marcus, R.M.~Westervelt, R.E.~Mirollo, Physica D {\bf 36}, 23 (1989).
%
\bibitem{economou}
E.N.~ Economou, {\it Green's Functions in Quantum Physics}, Springer Series in Solid 
State Physics, Vol. 7 Springer-Verlag, Berlin, (1979).
%
\bibitem{UMS98} A.V.~Ustonov, B.A.~Malomed, and S.~Sakai, Phys. Rev. B {\bf 57},
11691 (1998); O.M.~Braun and Yu.S.~ Kivshar, Phys. Rep. {\bf 306}, 1 (1998);
E.~Nylund, K.~Lindenberg, G.~Tsironis, J. Stat. Phys. {\bf 70}, 163 (1993).
%
\bibitem{HPC95} J.F.~Heagy, L.M.~Pecora, and T.L.~Carrol, Phys. Rev. Lett. {\bf 21}, 4185 (1995); 
J.F.~Heagy, T.L.~Carrol, and  L.M.~Pecora, Phys. Rev. E {\bf 50}, 1874 (1994).
%
\bibitem{FM96} L.M.~Floria and J.J.~Mazo, Adv. in Phys. {\bf 45}, 505 (1996).
%
\bibitem{gordon} A.V.~Ustinov, M.~Cirillo and B.A.~Malomed, Phys. Rev. B {\bf 47}, 8357 (1993).
%
\bibitem{pagano} S.~Pagano, M.P.~Soerensen, R.D.~Parmentier, P.L.~Christiansen, 
O.~Skovgaard, J.~Mygind, N.F.~Pedersen, and M.R.~Samuelsen, Phys. Rev. B {\bf 33}, 174 (1986).
%
\bibitem{BLD95} Y.~Braiman, J.F.~Lindner, and W.L.~Ditto, Nature {\bf 378}, 465 (1995).
%
\bibitem{braiman} Y.~Braiman, W.L.~Ditto, K.~Wiesenfeld, and M.L.~Spano, 
Phys. Lett. A {\bf 206}, 54 (1995).
%
\bibitem{GKKT98} A.~Gavrielides, Tsampikos Kottos, V.~Kovanis, and G.P~Tsironis, Phys. Rev. E 
{\bf 58}, 5529 (1998); Europhys. Lett. {\bf 44}, 559 (1998).
%
\bibitem{remark2} There is a variety of related quantities, that may be defined in a similar way. 
However, we use~$\sigma(jT)$ as a particular intuitive measure, that furthermore is easy to calculate.
%
\bibitem{WR90} H.G.~Winful and L.~Rahman, Phys. Rev. Lett. {\bf 65}, 1575 (1990).
%
\end{thebibliography}
\end{document}